\documentclass{article}
\usepackage{spconf,amsmath,graphicx}
\usepackage{amsmath,amsfonts}
\usepackage{algorithmic}
\usepackage{array}
\usepackage[caption,font=normalsize,labelfont=sf,textfont=sf]{subfig}
\usepackage{textcomp}
\usepackage{stfloats}
\usepackage{url}
\usepackage{verbatim}
\usepackage{graphicx}
\usepackage[singlelinecheck=off,justification=centering]{caption}
\usepackage{amsthm,amsmath,amssymb}
\usepackage{mathrsfs}
\usepackage{bm}
\usepackage{multirow}
\usepackage{booktabs}
\usepackage[numbers,sort&compress]{natbib}

\title{Scalable Multi-task Semantic Communication System with Feature Importance Ranking}
\name{Jiangjing Hu$^{\star}$, Fengyu Wang$^{\star}$$^{\ast}$, Wenjun Xu$^{\star}$$^{\dag}$, Hui Gao$^{\star}$, Ping Zhang$^{\star}$$^{\dag}$\thanks{This work was supported in part by the Fundamental Research Funds for the Central Universities under Grant 2022RC18 and in part by the National Natural Science Foundation of China under Grant 62293485.}
\address{$^{\star}$ Beijing University of Posts and Telecommunications, Beijing 100876, China
\\$^{\dag}$ Department of Mathematics and Theories, Peng Cheng Laboratory, Shenzhen 518066, China
\\$^{\ast}$ Corresponding author
 \\Email: \{Hujiangjing, fengyu.wang, wjxu, huigao, pzhang\}@bupt.edu.cn}}
\begin{document}
%
\maketitle
\begin{abstract}
Semantic communications are expected to be an innovative solution to the emerging intelligent applications in the era of connected intelligence. In this paper, a novel scalable multi-task semantic communication system with feature importance ranking (SMSC-FIR) is explored. Firstly, the multi-task correlations are investigated by a joint semantic encoder to extract relevant features. Then, a new scalable coding method is proposed based on feature importance ranking, which dynamically adjusts the coding rate and guarantees that important features for semantic tasks are transmitted with higher priority. Simulation results show that SMSC-FIR achieves performance gain w.r.t. individual intelligent tasks, especially in the low SNR regime.
\end{abstract}
\section{Introduction}
\label{sec:intro}

With the rapid increase of multimedia applications in the era of connected intelligence, unprecedented amounts of data need to be transmitted to serve various intelligent tasks, posing significant challenges to the existing communication systems. Semantic communications aim to extract and transmit concise semantics that are relevant to the tasks based on deep learning (DL), and are expected to be a promising solution to the next generation of communications.

Most existing works extract and transmit semantic information that is relevant to a specific task based on joint source-channel coding (JSCC). Particularly, in \cite{ref3,ref4}, JSCC is developed to encode and reconstruct text information. Jankowski $\emph{et}$ $\emph{al}$. propose two semantic frameworks for vehicle re-identification (ReID) tasks \cite{ref6}, which improve the ReID accuracy while reducing data volume evidently. A multi-user semantic extraction framework is further proposed to exploit the correlations among cameras for cooperative object identification in \cite{ref5}. However, the above systems are designed based on a single task, whereas most intelligent systems need to perform multiple tasks simultaneously. Therefore, parallel architectures are required for the multi-task scenario, increasing the network complexity exponentially.

Moreover, most existing systems are trained with a fixed coding rate \cite{+1,+2,vtm,semaudio}, which cannot adapt to the channel state information (CSI) and various task demands. To enhance the flexibility of semantic transmission, a variable length coding method is proposed in semantic communication systems, where the signal-to-noise ratio (SNR) is included during training to generate a MaskLayer to adjust the coding rate. However, extra complexity is caused by the independent network for MaskLayer training. Besides, the semantic features to be transmitted are selected without considering importance levels \cite{ref7,ref8,ref9,+3}, and thus, the important semantic features are probable to be abandoned, degrading the task performance.

In this paper, we propose a novel scalable multi-task semantic communication system. First, a joint semantic encoder is proposed for multi-task semantic extraction. Due to correlations among intelligent tasks, the extracted semantics of different tasks can be relevant, making the multi-task semantic features extraction more robust and efficient than parallel single-task coding. Furthermore, to improve the flexibility of feature transmission under dynamic channel conditions and varying task demands, a scalable coding method based on feature importance ranking (FIR) is proposed, where the extracted semantic features are ranked by the importance level, and then transmitted with different priorities. Numerical results show that the proposed system achieves scalable semantic information compression, and outperforms the traditional method and the-state-of-art DL-based scalable coding methods w.r.t. task performance for each task.

\section{SYSTEM MODEL}
\label{sec:format}

In this section, the system model is introduced. Specifically, a multi-task semantic communication system is considered, where $K$ intelligent tasks are executed simultaneously. The inputs of the system are assumed to be images, and the semantic features are extracted and then  transmitted to the receiver after scalable selection according to the CSI.
\begin{figure*}[htb]
  \centering
  \centerline{\includegraphics[width=14cm]{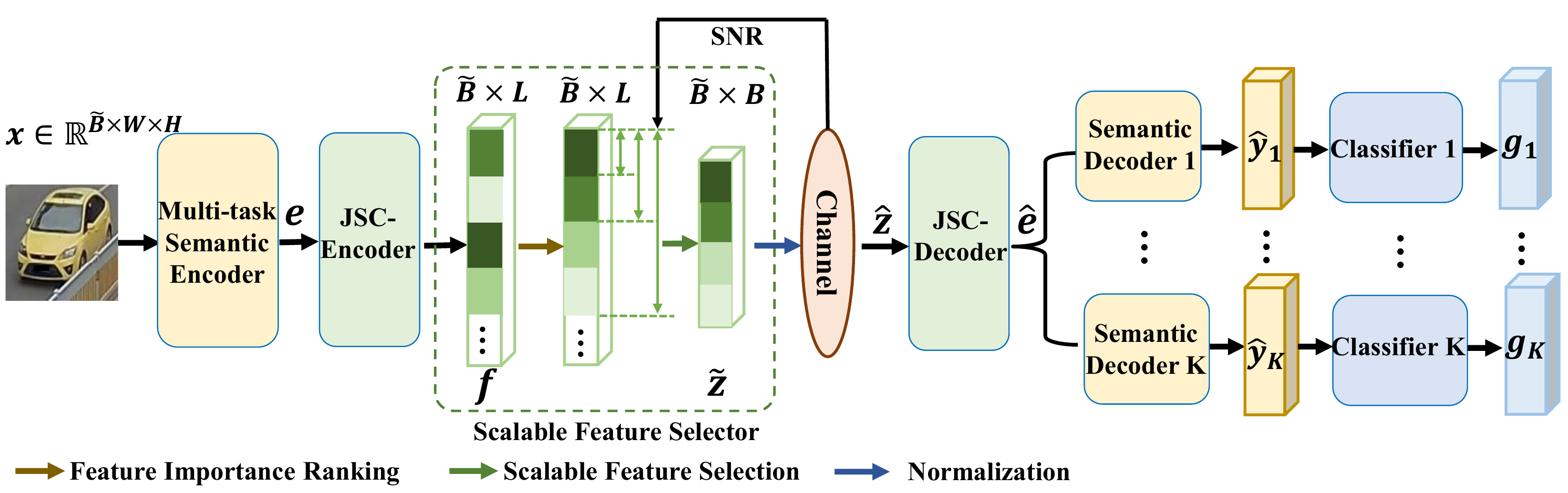}}
\captionsetup[figure]{labelsep=space}
\caption{The framework of the proposed SMSC-FIR. The different color depth of the features in $\bm{f}$ represents the corresponding importance level. }
\label{fig1}
\end{figure*}

As shown in Fig. \ref{fig1}, the transmitter consists of a multi-task semantic encoder, a joint source-channel encoder (JSC-Encoder), and a scalable feature selector, while the receiver consists of a joint source-channel decoder (JSC-Decoder) and multiple semantic decoders corresponding to different tasks. We denote the input as $\bm{x}\in\mathbb{R}^{\widetilde{B} \times W \times H}$, where $\widetilde{B}$ is the batch size, and $W$, $H$ is the width and height of the images. The semantic features are extracted as $\bm{e}\in\mathbb{R}^N$ by the semantic encoder, where $N$ is the number of features in $\bm{e}$. The JSC-Encoder encodes $\bm{e}$ as $\bm{f}\in\mathbb{C}^L$, where $L$ is the length of the feature vector. The feature vector selected by the scalable feature selector is $\bm{\widetilde{z}}\in\mathbb{C}^B$, where $B$ is the number of features to be transmitted according to CSI. Moreover, the feature importance ranking scheme is applied to $\bm{f}$ to select the features that are most relevant to the tasks (i.e., features with higher priority). The average power control is utilized to normalize the input signals of the channel, which is shown as $\bm{z}= \sqrt{PB}\frac{\bm{\widetilde{z}}}{||\bm{\widetilde{z}}||_2}$, where $P$ is the average power.

The channel is assumed to be Rician fading. The channel output can be described as $\bm{\hat{z}}=\bm{h}\bm{z}+\bm{n}$, where $\bm{n}\in \mathbb{C}^B$ denotes the white Gaussian noise, the value of which follows $\bm{n}_i\sim\mathcal{C}\mathcal{N}(0,\sigma^2)$. $\bm{h}$ represents the fading coefficient of the channel, which follows $\mathcal{C}\mathcal{N}(\sqrt{r/(r+1)},1/(r+1))$, where $r$ is set to 2.

At the receiver, the received semantic feature vector is decoded as $\bm{\hat{e}}\in\mathbb{R}^N$ by the JSC-Decoder. After multiple semantic decoders, the vector is restored to $[\bm{\hat{y}}_1,\bm{\hat{y}}_2,...,\bm{\hat{y}}_K]$ for the recognition of multi-tasks by the classifiers, where $K$ represents the number of multi-tasks.

\section{SCALABLE MULTI-TASK CODING WITH FEATURE IMPORTANCE RANKING}
\label{sec:pagestyle}
In this section, a scalable coding method based on feature importance ranking is proposed. First, a mapping scheme is designed to determine the length of transmission vector according to the CSI. Then, a feature importance ranking method is proposed to identify important features in the feature selector.
\subsection{Feature Quantity Mapping}
\noindent

We use the SNR to calculate the amount of information that can be transmitted under the given transmission bandwidth in a single time slot. On this basis, the number of features is determined. Specifically, the number of bits that can be transmitted is $V= TW\log_2(1+\rm{SNR})$, where $T$ is the transmission time slot, and $W$ is the channel bandwidth. Note that the semantic features are all double-precision floating-point numbers, the number of bits used by each feature is 64 bits. Then, the number of bits $V$ can be converted to the maximum number of semantic features $B$ by dividing by 64. The mapping scheme is applied to the feature selector at each time slot so that the transmitted feature vector length can be scalably changed with the channel conditions.
\subsection{Feature Importance Ranking}
\noindent

\begin{figure}[htb]
  \centering
  \centerline{\includegraphics[width=5.8cm]{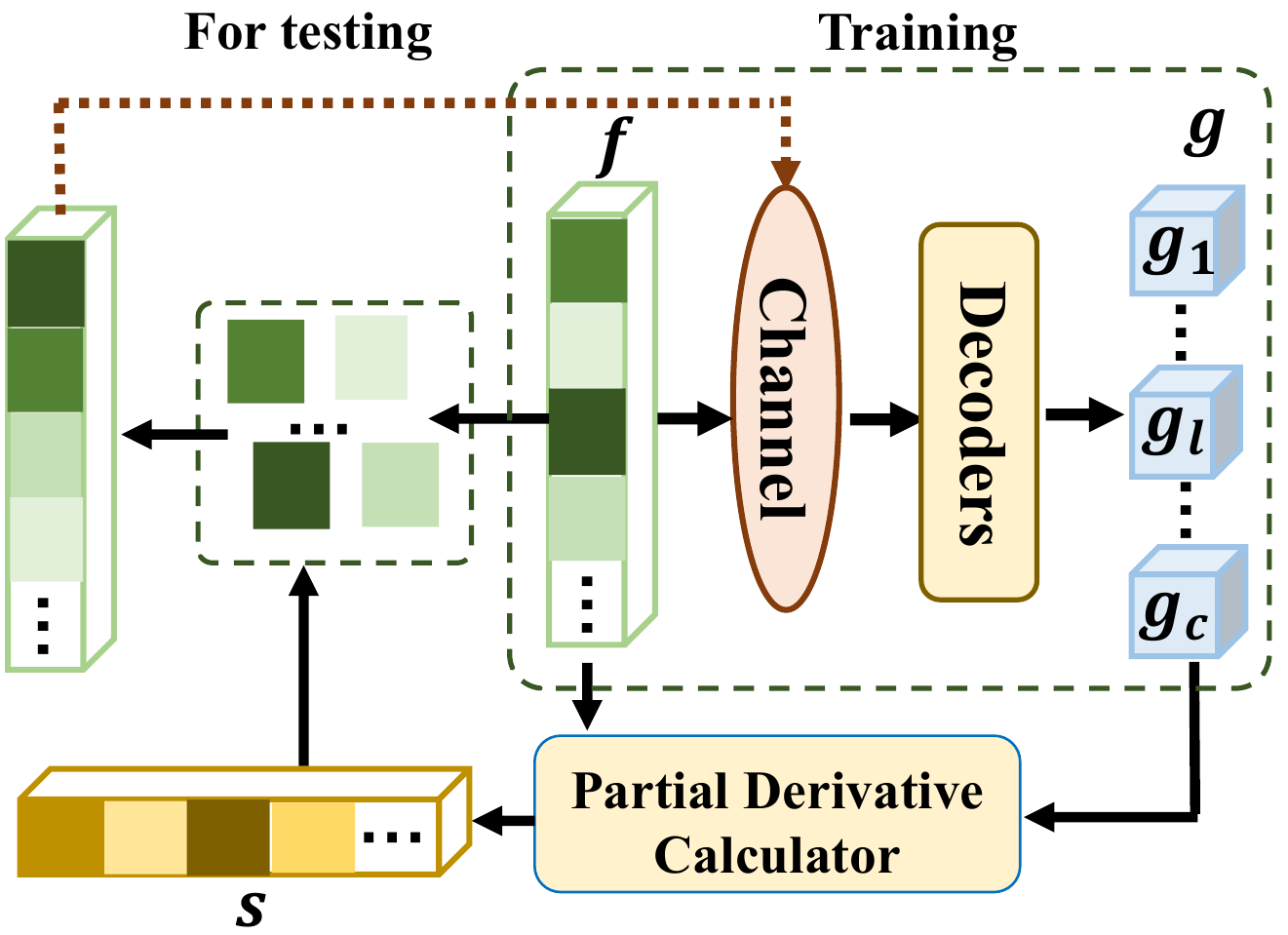}}
\caption{The framework of feature importance ranking.}
\captionsetup[figure]{labelsep=space}
\label{fig2}
\end{figure}
To identify the importance level of different features in $\bm{f}$ w.r.t. the specific task, we use Grad-CAM \cite{ref15,ref16}, which utilizes the class-specific gradient information to generate a heat-map to show the main concentrated regions of a CNN. For example, as shown in Fig. \ref{fig2}, the classification probability of each category of a specific task is obtained. The probability vector output by the classifier is denoted as $\bm{g}=[{g}_1,...,{g}_l,...,{g}_c]$, where $c$ indicates the total number of categories. The partial derivative feature vector $\bm{f}_l$ is obtained by calculating the gradient of ${g}_l$ relative to $\bm{f}$ as follows,
\begin{equation}
\label{deqn_ex1}
{\bm{f}_l}=\frac{\partial {{g}_l} }{ \partial{\bm{f}}}.
\end{equation}
The gradient information indicates the sensitivity of the feature vector $\bm{f}$ to the task outcome. The larger the gradient is, the more sensitive the feature is, indicating that the position is more significant to the category. $\bm{f}_l$ serves as the task sensitivity vector with the same dimension as $\bm{f}$, representing the sensitivity of each feature of $\bm{f}$.

The task sensitivity vectors of all the $K$ tasks are weighted on different tasks, then a feature importance measure vector with the same dimension as the original feature vector is obtained,
\begin{equation}
\label{deqn_ex1}
\bm{s}=\lambda_1\bm{f}_{l1}+...+\lambda_i\bm{f}_{li}+...+\lambda_K\bm{f}_{lK},
\end{equation}
where $\lambda_i$ denotes the weight of the $i$-th task, $\bm{f}_{li}$ is the task sensitivity vector corresponding to the $i$-th task. To the end, the features in $\bm{f}$ are selected based on $\bm{s}$ for various channel conditions, where the features corresponding to the positions with larger values in $\bm{s}$ are with higher priority.

\subsection{Training Strategy}
\noindent

To improve the efficiency of SMSC-FIR training, an E2E training strategy is proposed. First, the system is trained without the constraint of communication capacity, so that the feature importance measure vector $\bm{s}$ defined in Section 3.2 is obtained. DenseNet-121 pretrained on ImageNet \cite{ref13} is initialized to extract multi-task semantic features. In this paper, we focus on three representative tasks in intelligent traffic scenes for experiments, namely vehicle ReID, vehicle color classification, and vehicle type classification. The overall E2E training loss $L_{E2E}$ can be expressed as,
\begin{equation}
\label{deqn_ex1}
L_{E2E}=L_{T}+L_{CH},
\end{equation}
where $L_{T}$ and $L_{CH}$ denote the task loss and the channel transmission loss, respectively.

Specifically, $L_{T}$ can be calculated as,
\begin{equation}
\label{deqn_ex1}
L_{T}=\lambda_{r}L_{r}+\lambda_{c}L_{c}+\lambda_{t}L_{t},
\end{equation}
where $L_{r}$, $L_{c}$, $L_{t}$ are the loss of ReID, color classification, type classification, respectively. $\lambda_{r}$, $\lambda_{c}$, and $\lambda_{t}$ denote the weight of the corresponding task, respectively. Specifically, $L_{r}$ consists of the hard-mining triplet loss \cite{ref17} and the cross-entropy loss, which is defined as $L_{r}=\mathcal{L}_{ht}(a,p,n)+\mathcal{L}_{ce}(\bm{y},\bm{\hat{y}})$. $\mathcal{L}_{ht}(a,p,n)$ is the hard-mining triplet loss, where $a,p$ and $n$ represent anchor, positive and negative elements, respectively. $\mathcal{L}_{ht}(a,p,n)$ is defined as
\begin{equation}
\label{deqn_ex1}
\mathcal{L}_{ht}(a,p,n)= \max{\{(\max(d_{ap})-\min(d_{an})+\alpha),0\}},
\end{equation}
where $d_{ap}$ and $d_{an}$ are the distance between the extracted features of the anchor and positive/negative image. $\mathcal{L}_{ce}(\bm{y},\bm{\hat{y}})$ indicates the cross-entropy loss, which is defined as
\begin{equation}
\label{deqn_ex1}
\mathcal{L}_{ce}(\bm{y},\bm{\hat{y}})=\sum_{k=1}^Ky_k\log({\hat{y}}_k)+(1-{y}_k)\log(1-{\hat{y}}_k),
\end{equation}
where $\bm{y}$ denotes the label of the task and $\bm{\hat{y}}$ is the predicted one. For the other two classification tasks, the cross-entropy loss is also employed.

As for the channel transmission loss, $L_{CH}$ is measured by the mean square error (MSE),
\begin{equation}
\label{deqn_ex1}
L_{CH}=\frac{1}{N}\sum_{n=1}^N[||{e}_n-{\hat{e}}_n||^2_2],
\end{equation}
where $\bm{e}$ denotes the feature vector extracted by the semantic encoder, and $\bm{\hat{e}}$ means the feature vector restored by the JSC-Decoder.

Then, to obtain the scalable communication system with feature importance ranking, the SMSC-FIR is further retrained with limited bandwidth and time slot. The most important $B$ features are transmitted, where $B$ is determined by channel conditions.

\section{EXPERIMENTS AND RESULTS}
\label{sec:typestyle}

\begin{figure*}[htb]

\begin{minipage}[b]{0.33\linewidth}
  \centering
  \centerline{\includegraphics[width=6cm]{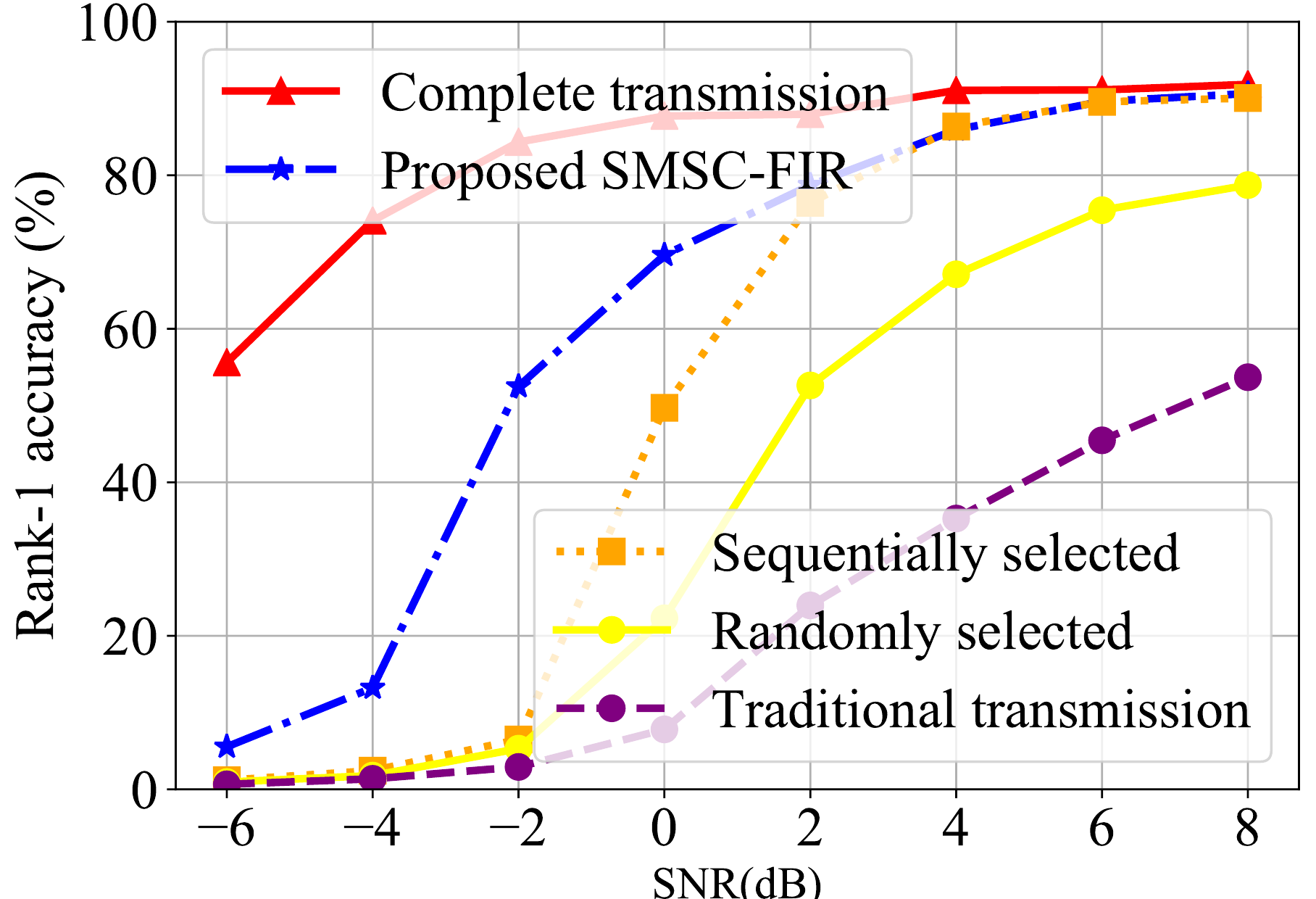}}
  \centerline{(a) ReID accuracy}\medskip
\end{minipage}
\begin{minipage}[b]{0.33\linewidth}
  \centering
  \centerline{\includegraphics[width=6cm]{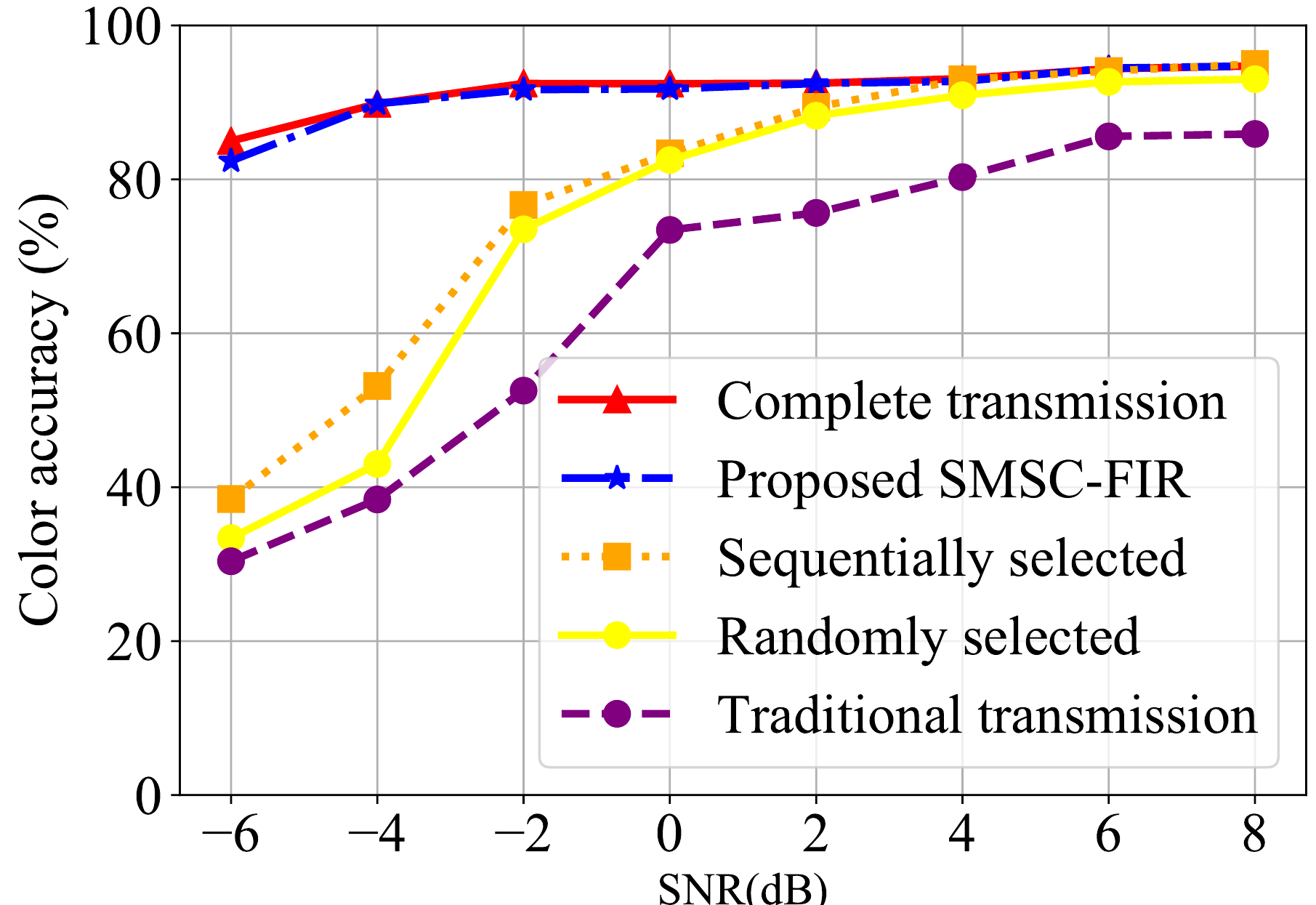}}
  \centerline{(b) Color accuracy}\medskip
\end{minipage}
\hfill
\begin{minipage}[b]{0.33\linewidth}
  \centering
  \centerline{\includegraphics[width=6cm]{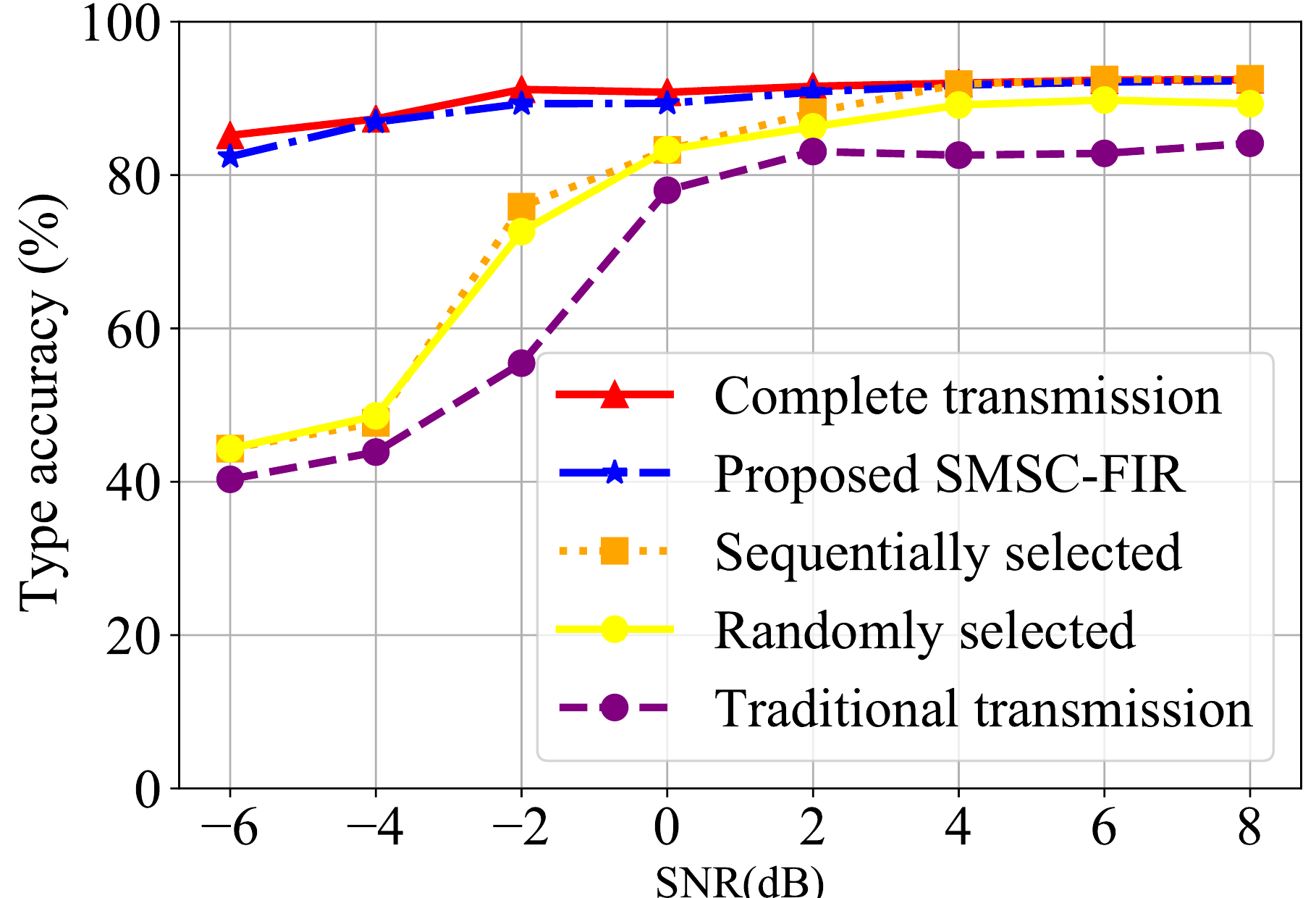}}
  \centerline{(c) Type accuracy}\medskip
\end{minipage}
\caption{Task accuracy, w.r.t. SNR}
\label{fig3}
\end{figure*}

In this section, simulations are performed to evaluate the proposed system. VeRi-776\footnote{VeRi-776 is captured by 20 cameras in urban areas, and consists of more than 50000 images of 776 vehicles. Moreover, there are 1678 query images, 11579 testing images, and 37778 training images, which are labeled with different attributes, such as type, color, and brand.}, a large-scale benchmark dataset for vehicle recognition \cite{ref14}, is leveraged for the multi-task learning model \cite{ref11}. The proposed system is compared with the state-of-the-art benchmarks in terms of the performance metrics of each task. The details are as follows:

\begin{itemize}
		\item[$\bullet$] {Traditional transmission method: JPEG, LDPC with $\frac{3}{4}$ rate and BPSK are used for source coding, channel coding and modulation, respectively. Then, the three intelligent tasks are performed using DenseNet based on the reconstructed images.}
		\item[$\bullet$] {DL-based scalable coding baselines: 1) Randomly selected method: The $B$ features to be transmitted are randomly selected from $\bm{f}$. 2) Sequentially selected method: The first $B$ features are selected and transmitted.
                          }
\end{itemize}

Note that JSCC is applied to the randomly selected and sequentially selected methods to achieve the best results. Moreover, the same training strategy as SMSC-FIR is adopted.

\subsection{Implementation Setup}
\noindent

The multi-task scalable communication system is evaluated under several SNR levels and the average power budget P is set to 1. During training, DenseNet121 is utilized as our backbone CNN, which is optimized by Adam \cite{ref19}. $\lambda_{r}$, $\lambda_{c}$, $\lambda_{t}$ are set to 1, 0.125, 0.125, respectively. The input images are first resized to $256 \times 256$, and the batch size is set to 32. The learning rate is initialized as 3e$-4$, and learning rate decay is applied. To illustrate the effectiveness of the joint multi-task semantic coding (MTC), we compare the task performance with the single-task coding (STC) methods by training three parallel JSCC architectures separately.

\subsection{Simulation Results}
\noindent

The recognition accuracy for each task of MTC and STC are presented in Tab .\ref{tab1} regarding rank-1 accuracy of ReID, average accuracy w.r.t. color and type classification. Both MTC and STC perform scalable transmission w.r.t. various SNR levels. As shown in Tab .\ref{tab1}, we can identify that the MTC method improves the performance for all three tasks. The reason is that the jointly extracted semantics can be relevant due to the correlations among different tasks, making the multi-task semantic feature extraction more robust and efficient than parallel single-task coding.

The multi-task performance w.r.t. different scalable methods are illustrated in Fig .\ref{fig3}. The complete transmission method serves as the upperbound, where the times of transfers are not limited. It is obvious that the DL-based scalable coding methods perform better than the traditional method, indicating the effectiveness of semantic transmission. The proposed SMSC-FIR achieves the best performance for all the three tasks compared with benchmarks. Specifically, at 0dB, in terms of ReID, SMSC-FIR improves the performance by 40.0$\%$ and 212.2$\%$, compared with the sequentially selected and randomly selected solutions, respectively. For color classification, the performance is improved by 11.0$\%$ and 9.8$\%$, and the performance is improved by 6.9$\%$ and 7.0$\%$ for type classification, respectively. The reason is that in SMSC-FIR, the most important features are used, due to the dedicated design of the FIR.

Furthermore, as shown in Fig .\ref{fig3}, for the proposed SMSC-FIR, the performance of classification w.r.t. color and type approaches the upperbound (a.k.a. complete transmission) even at low SNRs, illustrating the effectiveness of FIR. Note that different from the two classification tasks, the ReID performance is worse than the upperbound when SNR is low, and this is because more features are required in ReID. However, SMSC-FIR still achieves the best performance among all the benchmarks. Simulations shown in Fig .\ref{fig3} indicate that SMSC-FIR significantly improves the performance, especially in the low SNR regime.

\begin{table}[]

\scriptsize

\centering
 \resizebox{.99\columnwidth}{!}{

\begin{tabular}{cccccccccc}

\toprule

\multirow{2}{*}{SNR (dB)} & \multicolumn{3}{c}{Proposed MTC ($\%$)} & \multicolumn{3}{c}{STC ($\%$)}  \\

\cmidrule(r){2-4} \cmidrule(r){5-7}

&  Rank-1 acc.       &  Color acc.  &  Type acc.

&  Rank-1 acc.     &  Color acc.  &   Type acc.  \\

\midrule

$-6$             &5.54        & 82.40    & 82.40                   &0.83             & 64.55           & 46.95                   \\

$-4$             &13.23       & 89.83      & 86.92                   & 0.95           &83.51          &  85.81                   \\

$-2 $             &52.50       & 91.66    & 89.30                   &  5.13           & 88.83           &  88.25                    \\

$0$             &69.61    & 91.76      & 89.36                    & 48.15           & 90.84          & 88.19                      \\

$2$             &78.72        & 92.46        & 90.84                    & 66.69          & 91.81          & 88.21                   \\

$4$             &85.94          & 92.75     & 91.74                    & 77.83           & 92.73          & 89.13                      \\

$6$             &89.63      & 94.45        & 92.12                    & 84.62            & 94.40          & 92.10                      \\

$8$             &90.64       & 94.73       & 92.30                    & 84.68            & 94.67          & 92.24                     \\
\bottomrule

\end{tabular}}
\caption{Task accuracy of the proposed MTC method and the STC method w.r.t. SNR}
\label{tab1}

\end{table}

\section{CONCLUSION}
\label{sec:print}

In this paper, SMSC-FIR, a scalable multi-task semantic communication system with feature importance ranking, is proposed. The inter-task correlations are leveraged to obtain multi-task semantic features by a joint semantic encoder. Furthermore, to enable scalable transmission of features w.r.t. channel conditions, the importance level of each feature is evaluated, and the features with higher importance level are transmitted with higher priority. Simulation results show that the proposed SMSC-FIR outperforms the benchmarks, especially in the low SNR regime.

\clearpage
\bibliographystyle{IEEEtr}
\bibliography{reference}

\end{document}